\documentclass[a4paper,10.6pt]{article}
\usepackage[english]{babel}
\usepackage{amsfonts,amsmath,amssymb}
\usepackage{hyperref}
\usepackage{cite}
\numberwithin{equation}{section}
\usepackage{mathtools}
\usepackage{cleveref}
\crefmultiformat{equation}{(#2#1#3)}%
{ and~(#2#1#3)}{, (#2#1#3)}{ and~(#2#1#3)}
\crefrangeformat{equation}{(#3#1#4) --~(#5#2#6)}
\makeatletter
\renewcommand\@dotsep{8}   
\makeatother
\newcommand{\ch}{\mathcal{H}}
\newcommand{\eq}[1]{(\ref{#1})}

\allowdisplaybreaks[1]

\begin{document}

\thispagestyle{empty}  
\setcounter{page}{0}

\begin{center}
{\huge The effective field theory of multi-component fluids}\\
\vspace{1.2cm}
\textbf{Guillermo Ballesteros$^{a,b}$}, \textbf{Brando Bellazzini$^{c,d}$} and \textbf{Lorenzo Mercolli$^{e}$}
\vspace{0.5cm}

$^a$ \emph{Institut f\"{u}r Theoretische Physik, Universit\"{a}t Heidelberg, \\Philosophenweg 16, D-69120 Heidelberg, Germany}\\[3ex]

$^b$  \emph{University of Geneva, Department of Theoretical Physics \\
and Center for Astroparticle Physics (CAP),\\
24 quai E. Ansermet, CH-1211 Geneva 4, Switzerland }\\[3ex]

$^c$  \emph{Institut de Physique Th\'eorique, CEA-Saclay and CNRS URA-2306\\
F-91191 Gif-sur-Yvette Cedex, France}\\[3ex]

$^d$ \emph{Dipartimento di Fisica e Astronomia, Universit\`a di Padova and INFN Sezione di Padova\\ Via Marzolo 8, I-35131 Padova, Italy}\\[3ex]

$^e$  \emph{Department of Astrophysical Sciences, Princeton University, \\ Princeton, NJ 08544, USA }\\
\vspace{1cm}
\today\\
\vspace{1cm}

\textbf{Abstract}\\
\end{center}
\vspace{0.12cm}

We study the effective Lagrangian, at leading order in derivatives, that describes the propagation of  density and metric  fluctuations  in a fluid composed by an arbitrary number of interacting  components. Our results can be applied to any situation in cosmology where different species have non-gravitational interactions.

In time-dependent backgrounds, such as FLRW, the quadratic action can only be diagonalized at fixed time slices and flavour mixing is unavoidable as time flows. In spite of this, the fluid can be interpreted at the level of the linear equations of motion as an ensemble of individual interacting species. We show that interactions lead to anisotropic stresses that depend on the mixing terms in the quadratic action for perturbations. 

In addition to the standard entrainment among different components, we find a new operator in the effective action that behaves as a cosmological constant when it dominates the dynamics. 

\vfill

\noindent\rule{2cm}{0.4pt}
\begin{enumerate}{\small
\item[]\texttt{g.ballesteros@thphys.uni-heidelberg.de}
\vspace{-0.25cm}
\item[]\texttt{brando.bellazzini@cea.fr}
\vspace{-0.25cm}
\item[]\texttt{mercolli@astro.princeton.edu}}
\end{enumerate}

\newpage

\setlength{\textheight}{23.5cm}

\tableofcontents

\section{Introduction}

In cosmology we often encounter situations in which we have to describe several species simultaneously. A prototypical example  is the  $\Lambda$CDM model, according to which (and the latest data) the present universe contains a cosmological constant ($\sim 68\%$), cold dark matter ($\sim 27\%$), baryonic matter ($\sim 5\%$) and smaller amounts of radiation (and neutrinos) \cite{Ade:2013zuv}. In the standard picture, during the progressive cooling of the universe after reheating, some of these components ceased to interact among themselves and each of them contributes today to the Einstein equations with the energy-momentum tensor of a perfect fluid that is separately conserved. In more complicated models, some cosmological species are still allowed to interact at late times. For instance, there is considerable interest in constraining \cite{Amendola:2002bs,Amendola:2004ew,Valiviita:2009nu,Honorez:2010rr,DeBernardis:2011iw,Salvatelli:2013wra,Pettorino:2013oxa} possible interactions between the dark matter and a dark energy component that would be responsible for the accelerated expansion of the universe (e.g. \cite{Amendola:1999er,Mangano:2002gg,Comelli:2003cv,Farrar:2003uw,Pourtsidou:2013nha}). These interactions are typically characterized in an ad hoc manner, and it would be convenient to have a guiding principle that allowed us to study them within a general framework.

In additon, given the large number of models of the various epochs of the universe \cite{Martin:2013tda, Tsujikawa:2010sc,amendola2010dark,Clifton:2011jh,Yoo:2012ug}, confronting all the interesting proposals with the data is a daunting task. The fundamental tool to interpret the great majority of the observations is the theory of cosmological perturbations.  The reciprocal effects of the fluctuations pertaining to different species are very often modelled with (usually linearized) hydrodynamical equations, which depend on phenomenological coefficients that are then constrained  fitting the data\footnote{These equations (see e.g. \cite{Ma:1995ey}) are indeed a core part of the Boltzmann codes that are used to interpret the CMB data.}. In many models (for example of the late time universe)  the equations of scalar (or even vector) fields can also be written using fluid variables such as density, pressure and anisotropic stress. This adds up to the need for a general framework to classify and describe fluid interactions in cosmological perturbation theory. 

To simplify the process of model testing, a useful strategy is to search in the data for specific features shared by classes of models. In this work we develop a framework, exclusively based on symmetry arguments, that describes th
e dynamics of cosmological perturbations at large scales in multi-component fluids. We use the Effective Field Theory (EFT) of Fluids \cite{Dubovsky:2005xd, Endlich:2010hf,Ballesteros:2012kv} to describe the propagation of gapless sound waves, i.e phonons, in continuous media at low energies (or equivalently large distances). The power of the effective field theory is that given the low-energy degrees of freedom and the symmetries that characterize them, the form of the action is completely determined, with strong relations between its various terms.  This framework is then able to capture at once several microscopic models that share the same degrees of freedom and symmetry at large distances.

The EFT framework was used in  \cite{Ballesteros:2012kv}  to study of cosmological perturbations in single component perfect fluids. Our aim now is the description of fluids in cosmology that contain several components. Specifically, we are interested in determining  in the most general way the form of the allowed interactions between different cosmological species at the level of perturbations. The only  assumptions that we make are: 1) that the source of the energy-momentum tensor of the universe can be described in the continuum limit of a coarse graining at sufficiently large scales; 2) that the resulting system is symmetric under specific  transformations with a clear geometric meaning; and 3) that the only propagating degrees of freedom at low energies are associated to the spontaneous breaking of these symmetries.

Since we are interested in the long distance dynamics of sound waves, we can write an effective Lagrangian as a perturbative expansion in terms of derivatives. The relevant degrees of freedom in this series are the Goldstone bosons of the broken spacetime symmetries\footnote{This is the same basic principle that is used in the EFT of inflation \cite{Cheung:2007st} and dark energy \cite{Gubitosi:2012hu}.}. More concretely, the fluid backgrounds break space and internal translations spontaneously,  giving rise to  three Goldstone bosons per each component. These Goldstone bosons restore non-linearly general covariance and all the broken internal symmetries. The phonons are then represented by the displacement field $\pi$ with respect to the comoving coordinates of the fluid. Since the symmetries are non-linearly realized on $\pi$, there are stringent constraints on the couplings. This is specially useful to characterize the dynamics of sound waves in the extension of the EFT to multi-component fluids. 

The set of symmetries that is imposed to build the action of the continuum medium determines completely its physical behaviour. We will assume that each component of the multi-component fluid must be internally homogeneous and isotropic. Actually, we will impose an even stronger condition: invariance under volume preserving diffeorphisms, that leads to a perfect (and adiabatic) fluid in the single-component case. When there are several components, this symmetry does not produce a perfect fluid, due to the interaction between the components. Other symmetry choices are possible as well but, as we will see, the one we take already leads to a
rich phenomenology.  In particular, in addition to the known invariant operators built with two different four-velocities, we show that a new operator that can lead to $w=-1$ appears when there are at least four components in the fluid.

Then, guided by the currently prevailing Cosmological Principle, we will obtain the phonon action by requiring that the background states of all the components coincide with each other, in agreement with homogeneity and isotropy.  

There are several questions that we try to address in this work. As we have explained, our main goal is describing in broad generality the dynamics of perturbations when there are several cosmological species involved, possibly interacting non-gravitationally among them. In such a system, the mere definition of an individual species is a subtle matter. In a general  interacting system, a multi-component fluid, we can only define a single gravitational energy-momentum tensor for the whole ensemble of components, but we want to understand if and how it is possible to distinguish between different species. This will lead us to define {\it flavours} in a multi-component fluid through the action for the sound waves. We shall see that in dynamical backgrounds such us the ones that are typically considered in cosmology, the mixing of the perturbations prevents a proper definition of flavour. We can define conserved flavours only for time scales much shorter than the age of the universe.  Nevertheless, at linear order in cosmological perturbations, it is always possible to interpret a multi-component fluid as a mixture of different interacting species. In general, there is no unique way of mapping the Goldstone bosons $\pi$ into fluid variables. We will see that this translates into a relation between couplings and anisotropic stresses. 

The layout of the paper is the following. In Section \ref{sfi} we review the EFT of a perfect fluid. In Section \ref{eftmc} we generalize the theory to include multiple interacting components. We identify the operators that contribute at leading order in derivatives and obtain the gravitational energy-momentum tensor of the system. We also mention some properties of the new operator, $\Psi$, that appears when there are four or more components. In Section \ref{cosmophonon} we use those results to discuss cosmological phonons in a FLRW background. First, in Section \ref{diago}, we present the quadratic action for the phonons, introduce the concept of flavour for a multi-component fluid and discuss the impossibility of diagonalizing the phonon action due to the time dependence of the background. Then, in Section \ref{taxo},  we obtain the equations that describe the propagation of phonons and explain how to define fluid variables which allow to interpret the equations in terms of the perturbations of a system of interacting species. Finally, we conclude in Section \ref{conc}. In the the Appendix \ref{vort} we discuss the conservation of vorticity and its relation to the equations of motion of transverse phonons. 

Although this work focuses in cosmology and covariant theories of gravity (specifically on General Relativity), the formalism can be applied in other physical contexts\footnote{For example, it has been applied to study the oscillations of superfluid neutron stars \cite{Comer:1999rs,Andersson:2002jd}.}. Related work on multi-component fluids can be found in \cite{Comer:1993zfa, Andersson:2006nr, Comer:2011ss}.

\section{\,\,The EFT of a perfect fluid} \label{sfi}
In this section we briefly review the EFT of a perfect fluid, which describes the  dynamics of (gapless) sound waves (phonons) in a continuous medium.  For further details on this formalism and its application to cosmological perturbation theory, we refer the reader to \cite{Endlich:2010hf,Ballesteros:2012kv,Endlich:2012pz}.  The EFT of perfect fluids, as presented here  and in \cite{Dubovsky:2005xd, Endlich:2012pz, Ballesteros:2012kv}, blends together the variational principle for fluids \cite{Taub:1954zz,Carterx}, the pull-back formalism (see e.g. \cite{Brown:1992kc, Comer:1993zfa}) and the effective theory of phonons \cite{Leutwyler:1993gf,Leutwyler:1996er}. The application of the pull-back formalism in cosmology is now being developed. In \cite{Blas:2012vn} it was used to consider the possibility of Lorentz violation in dark matter. In \cite{Endlich:2012pz} a model of inflation based on a `solid' was proposed. In  \cite{Pourtsidou:2013nha} the formalism was applied to couple a quintessence scalar field to dark matter.

The kinematics of a continuous medium is fully described by the position, at each time, of all its elements, i.e. their trajectories. Since the medium is continuous, the elements can be labelled by a set of three continuous real functions $\Phi^i(t,x^j)$ of space and time. The trajectory of an element identified by $\Phi^i$ is then given by $x^i(t,\Phi^j)$ and it is clear that the spacetime fields $\Phi^i$ correspond to the coordinates of the elements in a system of reference that is comoving with the medium. In consequence, their variation along the flow is zero, which allows to express the four-velocity of the system in terms of $\Phi^i$ via
\begin{align}
u^\mu \partial_\mu \Phi^i=0\,,\quad u^2=-1\,.
\end{align}
The solution of these constraints is 
\begin{align} \label{veloc}
u^\mu=-\frac{1}{6\,b\,\sqrt{-g}}\epsilon^{\mu\alpha\beta\gamma}\epsilon_{abc}\partial_\alpha\Phi^a\partial_\beta\Phi^b\partial_\gamma\Phi^c\,,
\end{align}
where 
\begin{align} \label{dettb}
b=\sqrt{\det B}\,,\quad B^{ij}=\partial_\mu\Phi^i\partial^\mu\Phi^j\,.
\end{align}
An advantage of expressing the fluid velocity in terms of three scalar degrees of freedom is that we do not need to vary the action with a Lagrange multiplier to enforce the constraints as in \cite{Brown:1992kc, Comer:1993zfa}.

A volume preserving internal spatial diffeomorphism $\mathcal{V}\text{Diff}$ is a transformation of the comoving coordinates whose Jacobian determinant is equal to 1:
\begin{equation}\label{spatialdiff}
\mathcal{V}\text{Diff}:\,\Phi^i \rightarrow f^i(\Phi)\,,\quad  \det \left( \frac{\partial f^i}{\partial \Phi^j} \right) = 1 \;.
\end{equation}
The word `internal' simply means that these spatial diffeomorphisms act on the comoving  coordinates. Notice that these transformations include translations and SO(3) rotations and, therefore, any continuum medium that is symmetric under \eq{spatialdiff} is homogeneous and isotropic. We will soon see that imposing the symmetry \eq{spatialdiff} implies that we deal with a perfect fluid. If instead we choose to impose only the less stringent conditions of homogeneity and isotropy, we will obtain an imperfect fluid (because it has anisotropic stress) that has been referred as `solid' in the literature \cite{Dubovsky:2005xd}.

We are interested in describing the dynamics of perturbations of the fluid with respect to a reference frame where the unperturbed fluid is at rest. Therefore, we write
\begin{align} \label{phonon}
\Phi^i(t,x^j)=x^i+\pi^i(t,x^j)\,,
\end{align}
where the fields  $\pi^i$ represent the phonons around the solution $\Phi^i=x^i$. This background is invariant under a combination of an internal translation $\Phi^i\rightarrow\Phi^i+c^i$ and a space translation $x^i\rightarrow x^i-c^i$. This symmetry ensures the homogeneity of the environment in which the phonons propagate. Analogously, the unbroken diagonal combination of internal and space rotations ensures its isotropy. The phonons $\pi^i$ are the Goldstone bosons associated to this spontaneous symmetry breaking pattern  and, therefore, their low-energy (i.e. long-distance) dynamics is given by a derivatively coupled EFT. 

The symmetries of the $\Phi$ fields are non-linearly realized on the effective action for the phonons $\pi$, constraining the form of their interactions. In a static spacetime  (e.g. Minkowski) these interactions are then given at the lowest derivative orders by a Lagrangian which contains just a few parameters at each order in $\pi$. In a non-static homogeneous spacetime, such as FLRW, those parameters become functions of time. 

Specifically, given the symmetry \eq{spatialdiff}, the action for the phonons at lowest order in derivatives is obtained inserting \eq{phonon} into
\begin{align} \label{loa}
S_m=\int d^4x\, \sqrt{-g}\, F(b)\,,
\end{align}
where $b$ is defined in \eq{dettb}. The fluid Lagrangian is an arbitrary function of $b$ because that is the only possible invariant under \eq{spatialdiff} that gives one (and just one)  derivative per phonon. This function, which is sometimes named {\it master function} \cite{Andersson:2006nr}, determines the thermodynamical properties of the system. 

In cosmology we are interested in the interactions between the matter and the metric perturbations. To describe these interactions we consider, in addition to \eq{loa}, the standard Einstein-Hilbert action of General Relativity. 

The gravitational energy-momentum tensor of \eq{loa} is then
\begin{align} \label{emts}
T_{\mu\nu}=-\frac{2}{\sqrt{-g}}\frac{\delta S_m}{\delta g^{\mu\nu}}=-b F_b\, u_\mu u_\nu+(F-b F_b)\, g_{\mu\nu}\,,
\end{align}
where $F_b= d F/d b$ and the four-velocity is given by \eq{veloc}. This is the energy-momentum tensor of a perfect fluid with rest frame four-velocity $u^\mu$, pressure $p=F-b F_b$, and density $\rho=-F$.  

All the dynamics of the phonons is encoded in the function $F(b)$ and its derivatives, evaluated on the background. For example, in the case of an unperturbed Minkowski metric, the quadratic phonon Lagrangian derived from \eq{loa} is
\begin{align} \label{hpert}
\mathcal{L_\pi}^{(2)}=-\frac{\bar b \bar F_b}{2}\left(\dot\pi_\perp^2+\dot\pi_\parallel^2-\frac{\bar b \bar F_{bb}}{\bar F_b}(\nabla\cdot\pi_\parallel)^2\right)\,,
\end{align}
where the overbars indicate quantities that are evaluated on the background and we have split the phonons in transverse and longitudinal modes 
\begin{equation} \label{comps}
\pi=\pi_\parallel +\pi_\perp\qquad \nabla\cdot \pi_\perp=0\qquad \nabla\times\pi_\parallel=0\,.
\end{equation}
As discussed in \cite{Ballesteros:2012kv}, the equations of motion for $\pi$ can be recast into the usual Euler and continuity equations for a perfect fluid with adiabatic sound speed.

The four-vector
\begin{align} \label{nofentro}
J^\mu=-b\, u^\mu
\end{align} 
satisfies $J^\mu_{\,\,\,;\mu}=0$ identically and it can be identified with the entropy current, provided that the temperature of the fluid is defined as $T=-F_b$\,. Then, the comoving entropy density is simply $J^\mu u_{\mu}=b$\,. 

\section{\,\,The EFT of multi-component fluids} \label{eftmc}

In cosmology, we often deal with several species at the same time. For instance, we may want to characterize the late time evolution of dark matter perturbations in the presence of some form of dark energy or we may be interested in describing a small interaction between baryons and dark matter. If the different species only interact with each other gravitationally, the extension of the formalism outlined in the previous section is straightforward \cite{Ballesteros:2012kv}. The total effective Lagrangian is just the sum of the Lagrangians of the separate fluids. However, in absence of a compelling physical reason or a symmetry argument that prevents direct interactions, the assumption that each fluid has a separately conserved energy-momentum tensor becomes just a prejudice. This is particularly important for systems such as dark matter -- dark energy, for which our knowledge of the underlying theory is still very limited. An agnostic approach to this type of problems is particularly valuable, and this is what the EFT framework allows us to do. 

In what follows we generalize the EFT of perfect fluids to the case where there are several components. Our aim is to describe a system of $N$ cosmological species using $N$ copies of three scalar fields: 
\begin{align}
\Phi^i_A\,,\quad  {A=1,\ldots,N}\,,\quad {i=1,\ldots,3}.
\end{align}
We will call each of these triads a {\it component}, and avoid to refer to them as `fluids' or `species' for reasons that will soon become clear. In analogy with the previous section, we could naively expect that each triad of $\Phi$ fields (labelled with a Latin capital letter) would serve us to represent the comoving coordinates of a particular species. Although this starting point  allows us to construct the theory, we will  find that the actual definition of a species in the EFT framework is more subtle. First of all, we will see in this section that there is no neat way in which the energy-momentum tensor of the system \eq{emt} can be chopped into pieces that we can associate to different species. 

Given this difficulty, we will try in Section \ref{diago} to define separate species using the quadratic action for the perturbations, introducing the concept of {\it flavour} as an independently propagating  excitation. We will find that the time dependence of the mixing terms that occurs in dynamical spacetimes makes impossible  the identification of conserved flavours at all times. Flavours that are identified in the quadratic phonon action at a certain time get eventually mixed through the time evolution of the system.  In spite of this flavour non-conservation, we will see in Section \ref{taxo} that we can actually identify individual species that are maintained along the time evolution at the level of the equations of motion. 
 
Let us now proceed with the construction of the EFT for a multi-component system. The first question that we must answer is which are the symmetries that characterize the theory. A natural generalization of \eq{spatialdiff} consists in imposing the invariance of the action under
\begin{equation}\label{difflavour}
\mathcal{V}\text{Diff}_A:\,\Phi^i_A \rightarrow f_A^i(\Phi_A)\,,\quad  \det \left( \frac{\partial f_A^i}{\partial \Phi_A^j} \right) = 1 \;,
\end{equation}
allowing for different diffeomorphisms $f_A$ to act on each component. The overall symmetry of the action would then be the direct product of $N$ copies of \eq{difflavour},  $\mathcal{V}\text{Diff}^N=\mathcal{V}\text{Diff}_{A_1}\times\ldots\times \mathcal{V}\text{Diff}_{A_N}$. For each of $\Phi_A$\,, we can  construct the analogous of the determinant \eq{dettb} as
\begin{align} \label{detta}
b_A=\sqrt{\det B_A}\,,\quad B_A^{ij}=\partial_\mu\Phi_A^i\partial^\mu\Phi_A^j\,,
\end{align}
which clearly is invariant under $\mathcal{V}\text{Diff}^N$. Another possibility consists in imposing a much weaker symmetry, $\Phi^i_A \rightarrow f^i(\Phi_A)\, \forall A$\,, where $f^i$ is the same for all the components.  This symmetry gives raise to a less symmetric situation, with more invariants contributing to the lowest order action for the phonons than in the case of $\mathcal{V}\text{Diff}^N$. For instance, given any pair of  fields, say $\Phi_A$ and $\Phi_B$, the determinant of the matrix $\partial_\mu\Phi_A^i\partial^\mu\Phi_B^j$ is invariant under a single element of $\mathcal{V}\text{Diff}$ acting on both $\Phi_A$ and $\Phi_B$, but not under a generic element of $\mathcal{V}\text{Diff}^N$. In this work we choose to focus exclusively in the larger  symmetry $\mathcal{V}\text{Diff}^N$, which is also the implicit assumption in \cite{Comer:1993zfa}. The physical meaning of our choice is that each component $\Phi_A$ can be relabelled  independently (with a volume preserving transformation)  without altering the physical properties of the whole system. In this sense, each $\Phi_A$ remains a separate entity from the others. 

The four-velocity \eq{veloc} and the current  \eq{nofentro} can be defined for each component in the obvious way: adding a subscript $A$ to the comoving coordinates $\Phi$ in those expressions. For instance, the four-velocity of the component $A$ would just be:
\begin{align} \label{rfa}
u_{A}^\mu\partial_\mu\Phi_A^i=0\,,\quad u_A^2=-1\,.
\end{align}
With the corresponding  currents
\begin{equation}
J^\mu_A = -b_A u^\mu_A\,,
\end{equation}
we introduce the following contractions: 
\begin{align} \label{cont}
J_{AB}=-g_{\mu\nu}\, J^\mu_{A}J^\nu_{B}\,,
\end{align}
which allow to express \eq{detta} as $b_A=\sqrt{J_{AA}}$\,. Since each current $J^\mu_A$ is invariant under $\mathcal{V}\text{Diff}_A$\,, the new scalars $J_{AB}$ are invariant under the symmetry $\mathcal{V}\text{Diff}^N$. 

In order to write the most general action for the phonons at the lowest order in derivatives that is symmetric under $\mathcal{V}\text{Diff}^N$, we need to identify all the  possible invariants that carry only one derivative per $\pi$. This is the case for $J_{AB}$, but this type of invariants do no exhaust  all the possibilities. With four or more components, the following invariants should be included as well:
\begin{align} \label{theta}
\Psi_{IABC}=\sqrt{-g}\,\epsilon_{\mu\alpha\beta\gamma}\,J^\mu_I\, J^\alpha_A\, J^\beta_B\, J^\gamma_C\,.
\end{align}
To the best of our knowledge, these invariants had not been considered until now. Since they are completely antisymmetric under the permutation of the components, it is  impossible to obtain a non-zero $\Psi_{IABC}$ if  $N<4$. If two currents, say $J_I^\mu$ and $J_A^\mu$, are parallel,  $\Psi_{IABC}$ is zero. In general, there exist  $\binom{N}{4}$ different $\Psi_{ABCD}$ invariants that should in principle be considered. However, it is actually enough to consider just one, say $\Psi_{1234}$, because any other $\Psi_{ABCD}$ can be expressed in terms of the $J_{AB}$ invariants and $\Psi_{1234}$. In four dimensions, we can choose a basis of four linearly independent currents $J_{I}^\mu$\,, $I=1\,\ldots\,4$ and express any current $J_A^\mu$ as a linear combination
\begin{align}
J_A^\mu \; = \; h_1(J_{KL}) \, J_1^\mu + h_2(J_{KL}) \, J_2^\mu + h_3(J_{KL}) \, J_3^\mu + h_4(J_{KL}) \, J_4^\mu \,,
\end{align}
where the functions $h_{1,2,3,4}$ depend on all possible scalar products of two currents (which is nothing but $J_{KL}$). If the currents of the chosen basis are orthogonal to each other, the functions $h_{1,2,3,4}$ depend only on a single $J_{KL}$, i.e. $h_{1} = J_{A1}/b_1$ etc. It is then clear that even if the four-velocities of the different components are not orthogonal, it is always true that $\Psi_{ABCD} = h(J_{KL})\Psi_{1234}$ for some function $h$.  From now on, we will therefore denote $\Psi\equiv\Psi_{1234}$.

Expanding $\Psi$ at leading order, we get
\begin{align}\label{psi}
\Psi=-\frac{\epsilon_{ijk}}{a^{12}}\left(\dot\pi^i_1\dot\pi^j_2\dot\pi^k_3-\dot\pi^i_1\dot\pi^j_2\dot\pi^k_4+\dot\pi^i_1\dot\pi^j_3\dot\pi^k_4
-\dot\pi^i_2\dot\pi^j_3\dot\pi^k_4\right)+\mathcal{O}(4)\,,
\end{align}
where we have taken a common background $x^i$ for all $\Phi_A$. This shows that $\Psi$ starts only at cubic order (while $J_{AB}$ contains quadratic terms). 

Having identified the invariants, we can write the most general action that gives the dynamics  of the phonons  (at lowest order in derivatives) in a multi-component fluid in General Relativity:
\begin{align} \label{mf}
S=\frac{1}{16\pi G}\int\, d^4x \, \sqrt{-g}\, R+\int \,d^4x\, \sqrt{-g}\, F(J_{AB}\,,\Psi)\,,\quad   A \leq B=1,\ldots, N\,,
\end{align}
where the inequality $A\leq B$ avoids redundant operators, since $J_{AB}=J_{BA}$. Therefore, for $N$ components, the action \eq{mf} is a functional of $\Psi$  and $N(N+1)/2$ different $J_{AB}$.   If the $N$ components only interact through gravity, we recover the results of \cite{Ballesteros:2012kv} and, in the case of a single component, this action reduces to \eq{loa}.

Following \cite{Andersson:2006nr} we call {\it entrainment} the dependence of the master function $F$ on $J_{AB}$ with $A\neq B$. This name suggests that the contraction of two non-parallel $J^\mu_A$ couples the modes of the different components. As we will later see, what actually happens is that unavoidable phonon coupling appears (even at quadratic order in the perturbations) also due to the $J_{AA}$ contributions, provided that the propagation takes place in a non-static spacetime. This makes the phonon coupling of different components a general feature in most relevant backgrounds for  cosmology. 

Since $F$ is an arbitrary function of its arguments, there is no way in which we can decompose the action \eq{mf} into a sum of actions for separate species plus an interaction term. The gravitational energy-momentum tensor of the system is covariantly conserved as a whole, and  we cannot define it for each of the components separately. From \eq{mf} we obtain
\begin{align} \label{emt}
T_{\mu\nu}=\left(F-3\,\Psi\, F_{\Psi}-2\sum_{\{AB\}}^N  F_{J_{AB}}\,J_{AB}\right)g_{\mu\nu}-2\sum_{\{AB\}}^N F_{J_{AB}}\,J^{A}_{(\mu}J^{B}_{\nu)}
\end{align}
where, as in \eq{emts}, the subscripts carried by the master function $F$ represent differentiation with respect to that operator, i.e. $F_\Psi=\partial F/\partial\Psi$, $F_{J_{AB}}=\partial F/\partial J_{AB}$. The sums extend over values of the indices that select inequivalent operators, so we use the following notation:  $\sum_{\{AB\}}^N =\sum_{A=1}^{N}\,\sum_{B=A}^N$\,. Besides, the indices inside a small parenthesis are symmetrized over. For example, taking the energy-momentum tensor: $2\,T_{(\mu\nu)}\equiv T_{\mu\nu}+T_{\nu\mu}$. 

As we have discussed, it would be wrong to say that the action \eq{mf} is an ensemble of fluids. What it really describes is a non-elementary fluid (formed by several components) whose energy-momentum tensor is given by \eq{emt}. This is the reason why we have avoided using the word fluid to refer to the fields $\Phi_A$ and instead we have called them components. 

In order to understand what type of fluid we have got with the symmetry $\mathcal{V}\text{Diff}^N$, we can project \eq{emt} in a frame to obtain the corresponding energy density, pressure, etc. Let us recall that in cosmology a frame is a four-velocity that can be thought of as characterizing the state of movement of a certain observer. Given a frame $\upsilon^\mu$, any energy-momentum tensor can be decomposed into
\begin{align} \label{emtg}
T^{\mu\nu}=\rho_{[\upsilon]}\, \upsilon^\mu \upsilon^\nu+p_{[\upsilon]}\left(g^{\mu\nu}+\upsilon^\mu \upsilon^\nu\right)+2q_{[\upsilon]}^{(\mu} \upsilon^{\nu)}+\pi_{[\upsilon]}^{\mu\nu}\,.
\end{align}
The bracket subscripts indicate that the quantities carrying them are obtained as projections of the tensor in the frame within the brackets and represent the properties of the fluid as seen by that observer. We would like to project \eq{emt} on its own energy (Landau-Lifshitz) frame $u^\mu_E$\,, given by the eigenvalue equation:
\begin{align} \label{rf}
T_{\mu\nu}u^{\nu}_E+\rho_{[E]}\, u_{\mu}^E=0\,,
\end{align}
which is the condition for vanishing energy flux $q^\mu$ (that is sometimes called heat flux). In the single component case \eq{emts}, the equation \eq{rf} gives precisely the rest frame velocity \eq{veloc}, with eigenvalue $\rho_{[E]}=-F$. In the multi-component case \eq{emt} that we are now dealing with, it is far from obvious how to solve (fully non-linearly)  the equation \eq{rf}. However, it is possible to find the solution order by order in perturbation theory, after a background solution of the equations of motion is chosen, as we will see in equations \eq{trf}--\eq{secdenp}.

In a  FLRW universe, the energy density and pressure are given by 
\begin{align} \label{dep}
\bar\rho=-\bar F\,,\qquad
\bar p=\bar F-2\,\bar b\,^2\sum_{\{AB\}}^N\bar{F}_{J_{AB}}\,,
\end{align}
where $\bar b=1/a^3$ and $a$ is the scale factor. These expressions can be read from the energy momentum tensor \eq{emt} taking into account  that any four-velocity in an exact FLRW has components $u^0=1/a$ and $u^i=0$ (using conformal time). We see that the invariants $J_{AB}$ contribute to the background pressure but not to the background density. Notice also that the equation of state $w$,  which is defined as the ratio between the background energy density and pressure is then
\begin{align} \label{eom}
1+w=\frac{2\,\bar b\,^2}{\bar F}\sum_{\{AB\}}^N \bar{F}_{J_{AB}}=2\sum_{\{AB\}}^N \overline{\frac{\partial \log F}{\partial\log J_{AB}}}\,.
\end{align}

Remarkably, the invariant $\Psi$ is zero in a FLRW background. In a general situation, when both $\Psi$ and $J_{AB}$ are important,  the operator $\Psi$ is invisible at the background level, but it has density and pressure perturbations (phonons) at cubic and higher orders in $\pi$.

However, in a model in which the operators $J_{AB}$ were negligible in comparison with $\Psi$, the equation of state would be exactly equal to $-1$. This  occurs regardless of the number of components, provided that there are at least four distinct ones. Actually, the energy-momentum tensor \eq{emt} tells us that $\rho+p$ would be zero for any metric, whenever the operator $\Psi$ dominates the dynamics. Let us suppose for a moment that the only important operator in the multi-component fluid is indeed $\Psi$. Then,  the energy-momentum tensor  is exactly
\begin{align} \label{EMpsi}
T_{\mu\nu}=\left(F-3\Psi F_\Psi\right)g_{\mu\nu}
\end{align}
Obviously, $T^{\mu\nu}_{\,\,\,\,\,\,\,;\nu}=0$ implies that $\partial_\mu\left(F-3\Psi F_\Psi\right)=0$ and therefore \eq{EMpsi} represents a cosmological constant, which takes the value
\begin{align}
\Lambda =-8\pi G\left(F-3\Psi F_{\Psi}\right)
\end{align}
in any frame. Given that $F$ is a function of $\Psi$, there are two possibilities to make this energy density, $\rho=\Lambda/8\pi G$, constant (in time and space). One of them is to impose that $\partial_\mu\Psi=0$. The other possibility consists in choosing the functional form of $F(\Psi)$ adequately. In particular, if we  take
\begin{align}
F=F_0+3\Psi_0F_{\Psi}|_{\Psi=\Psi_0}\left(\left(\frac{\Psi}{\Psi_0}\right)^{1/3}-1\right)\,,
\end{align}
where $F_{\Psi}|_{\Psi=\Psi_0}$ is the derivative of $F$ with respect to $\Psi$ at $\Psi=\Psi_0$, the energy density  satisfies $\partial_\mu\rho=0$ for any spacetime function $\Psi$. 

\section{\,\,Cosmological phonons in a multi-component fluid} \label{cosmophonon}
We want to study fluctuations around $\Phi^i_A=x^i$ in a perturbed FLRW universe in Poisson gauge:
\begin{align} \label{poisson}
ds^2=a^2\left(-(1+2\psi)d\tau^2+2\nu_id\tau d x^i+\left[(1-2\phi)\delta_{ij}+\chi_{ij}\right]dx^i dx^j\right)\,,
\end{align}
The background $\Phi^i_A=x^i$ is universal for all the components, which is consistent with the kind of spacetime that we want to describe.  At linear order in $\pi$, the operators \eq{theta} do not contribute and we just need to focus on  \eq{cont}. To find out the type of fluid that \eq{emt} is at this order, we study its properties on its own energy frame. As we mentioned in the previous section, we can easily compute the energy frame at first order, solving the equation \eq{rf}. The result is 
\begin{align} \label{trf}
u^0_{E}=\frac{1}{a}+\mathcal{O}(2)\,,\quad u^i_{E}=-\frac{\sum_{{\{AB}\}}\bar F_{J_{AB}}\left(\dot\pi^i_A+\dot\pi^i_B\right)}{2a\sum_{{\{AB}\}}\bar F_{J_{AB}}}+\mathcal{O}(2)\,,
\end{align}
which is nothing but a weighted sum\footnote{See the equation \eq{eom} for the equation of state of the fluid.} of the rest frames of the different components. The corresponding eigenvalue gives the density perturbation (defined as $\delta X=X-\bar X$) at linear order:
\begin{align} \label{firstdenp}
\delta\rho_{[E]} =-\,\bar b\,^2\sum_{\{AB\}}^N  \bar F_{J_{AB}} \left(6\phi+\nabla\cdot\pi_A+\nabla\cdot\pi_B\right)\,.
\end{align}
We then compute the pressure contracting the energy-momentum tensor with the projector on hypersurfaces orthogonal to the four-velocity \eq{trf}:
\begin{align}
\delta p_{[E]}=\delta\rho_{[E]}-2\,\bar b\,^4\sum_{\{AB\}}^N\sum_{\{IJ\}}^N  \bar F_{J_{AB}J_{IJ}} \left(6\phi+\nabla\cdot\pi_I+\nabla\cdot\pi_J\right)\label{secdenp}\,.
\end{align}
These expressions reduce  to the ones obtained in \cite{Ballesteros:2012kv} for a single component fluid. 

The standard  continuity and Euler (actually Navier-Stokes) equations at linear order for an arbitrary $T^{\mu\nu}$ in its energy frame are well-known:
\begin{align} \label{continuity}
\frac{\partial}{\partial\tau}\left(\delta\rho\right) &=-3\ch\left(\delta\rho+\delta p\right)+\left(\bar\rho+\bar p \right)\left(3\dot\phi-\theta\right)\\
\label{euler}
\bar\rho\,\frac{\partial}{\partial\tau}\Big((1+w)\,\theta\Big) &=(3w-1)\left(\,\bar\rho+\bar p\,\right)\ch\theta -\nabla^2 \delta p+\left(\,\bar\rho+\bar p\,\right)\nabla^2(\sigma-\psi)\,.
\end{align}
In these equations, $\ch$ is the conformal Hubble parameter, the scalar anisotropic stress is
\begin{align}
(\bar\rho+\bar p)\nabla^2\sigma=\sum_{i, j}\partial_i\partial_j T^i_j-\frac{1}{3}\sum_k\nabla^2T^k_k\,,
\end{align}
and the velocity divergence $\theta=\nabla\cdot v$ comes from the spatial part of the four velocity $u^i=a^{-1}(1-\psi+\ldots)v^i$. 
Therefore, the expression for $\theta$ in the energy frame \eq{trf} is just
\begin{align} \label{thetar}
(1+w)\bar\rho\,\theta_{[E]}=\bar b^2\sum_{\{AB\}}^N\bar F_{J_{AB}}\left(\nabla\cdot\dot\pi_A+\nabla\cdot\dot\pi_B\right)\,.
\end{align}
Using the expressions above, one can easily check that the continuity equation \eq{continuity} is identically satisfied. As explained in \cite{Ballesteros:2012kv} (for the case of a single component), once the rest frame four-velocity has been identified, the continuity equation is devoid of dynamical content. We see that the same occurs for a fluid that contains several interacting components. Furthermore, the Euler equation \eq{euler} holds provided that $\sigma$ is zero at linear order. This can be checked using the equations of motion of the longitudinal phonons \eq{complexeuler}, that we give in Subsection \ref{taxo}. A combination of these equations gives the Euler equation \eq{euler}. This means that the system described by the action \eq{mf}  behaves as a perfect fluid at this level. Indeed, taking the difference between \eq{emt} and \eq{emtg} (with $\upsilon^\mu=u^\mu_{E}$ and the above expressions for the pressure and density perturbations), it is straightforward to check that the anisotropic stress $\pi^{\mu\nu}_{[E]}$ is zero (in the energy frame of the fluid) and therefore the fluid is perfect (at linear order). 

Let us notice that from the expressions \eq{ede} and \eq{pre}, we can easily get the linear energy density and pressure perturbations  in any component frame $u^\mu_{A}$.
These perturbations turn out to be the same as \eq{firstdenp} and \eq{secdenp} and are therefore independent of the specific component on which we project. On the other hand, the energy flux in the frame $u^\mu_A$ is different from zero already at linear order, measuring the mismatch with respect to the energy frame of the fluid:
\begin{align}
q_0^{[A]}\sim\mathcal{O}(2)\,,\quad\,q_i^{[A]}=-\frac{\bar b\,^2}{a}\sum_{\{BC\}}^N \bar F_{J_{BC}}\left(\dot\pi^i_B+\dot\pi^i_C-2\dot\pi^i_A\right)+\ldots
\end{align}
One can check that the fluid also exhibits anisotropic stress when viewed in a frame comoving with one of the components.

\subsection{Quadratic action and flavour} \label{diago}
In this section we work with the quadratic action for the Goldstone bosons $\pi_A$. This action describes the propagation of sound waves in the multi-component fluid at lowest order in cosmological perturbations. In Section \ref{taxo}, we will use it to obtain the linear equations of motion for the phonons and interpret these equations in terms of standard fluid variables: density and pressure perturbations, etc. As we are going to see now, the $\pi_A$ fields of different components interact derivatively in the quadratic action and, in general, this cannot be avoided using field redefinitions. This property is a signature of the intrinsic interacting nature of the multi-component fluid and, even though it hinders the possibility of defining different flavours at the level of the action, we will see in the next section that it does not impede us from interpreting the multi-component fluid in terms of different (interacting) species. 

Let us now discuss the second order action for the phonons, including their interactions with the metric perturbations:
\begin{align}\nonumber
S^{(2)}=&\sum_{\{AB\}}\sum_{\{CD\}}\int d^4x\, \frac{\bar F_{J_{AB}J_{CD}}}{2a^8}\big(\, 6\phi\left(\nabla\cdot\pi_A+\nabla\cdot\pi_B+\nabla\cdot\pi_C+\nabla\cdot\pi_D\right)+\Pi_{AC}+\Pi_{AD}+\Pi_{BC}+\Pi_{BD}\big)\\+&\sum_{\{AB\}}\int d^4x\, \frac{\bar F_{J_{AB}}}{a^2}\big(-\dot\pi_A\cdot\dot\pi_B+\Pi_{AB} +\nu\cdot\left(\dot\pi_{A\perp}+\dot\pi_{B\perp}\right)+\left(3\phi+\psi\right)\left(\nabla\cdot\pi_A+\nabla\cdot\pi_B\right)\big)+\ldots\,, \label{seca}
\end{align}
where we use the following notation: $\Pi_{AB} \equiv\left(\nabla\cdot\pi_A\right)\left(\nabla\cdot\pi_B\right)$. The ellipsis in \eq{seca} indicates terms that involve only metric perturbations, which are not needed to obtain the equations of motion for the $\pi_A$. The operators $\Psi_{ABCD}$, defined in   \eq{theta}, play no role at this order in perturbations. They only start to become important in the cubic action, i.e. for the second order equations of motion. 

The most remarkable property of the phonon action \eq{seca} is the mixing between derivatives of different components. The mixing occurs on spatial and temporal derivatives separately. The two types of derivatives cannot mix with each other at quadratic order\footnote{However, space and time derivatives do mix at higher orders.} due to the index structure of the phonons, which comes from the pattern of symmetry breaking \eq{phonon}. In order to understand if the mixing is a real feature or just an effect of our choice of variables, we have to determine if the action \eq{seca} can be diagonalized (in such a way that only $(\dot\pi_A)^2$ and $(\nabla\cdot\pi_A)^2$ would appear on it after a linear field redefinition). If the diagonalization is possible, we would be able to write the phonon Lagrangian of the fluid as a sum of actions like \eq{hpert}, identifying clearly the individual propagating degrees of freedom, that we want to call {\it flavours}. Instead, if the action cannot be diagonalized, any $\pi_A$ will have a certain probability of oscillating spontaneously into a different $\pi_B$ after a given propagation time, very much in the same way that we know it occurs for neutrinos. 

There are several difficulties that complicate the diagonalization and hence the definition of flavours. First of all, the $\pi$ fields do not only mix among themselves but also with the scalar and vector metric perturbations. This mixing between metric and matter variables is gauge dependent. We have chosen to write the metric in Poisson gauge \eq{poisson}, but in any other gauge the mixing will look different. In fact, the phonons $\pi$ are gauge dependent quantities themselves, since they are defined with respect to a specific choice of coordinates $\tau$ and $x^i$. Flavours should be gauge independent; and we can expect to overcome this problem by writing the action \eq{seca} in terms of gauge invariant variables (that will contain both matter and metric perturbations). 

There is however a bigger hurdle in the way towards a full diagonalization of the action: the mixing matrices of both (spatial and temporal) parts of the kinetic term are time dependent. We define the flavours as the variables in which both mixing matrices are simultaneously diagonal. The problem is that the time dependence of these matrices makes this simultaneous diagonalization impossible by local (in time) field redefinitions. In a static spacetime, the mixing coefficients (and all the coefficients of the effective action) are constant numbers and it is easy to check that the action can be diagonalized. This works in the Minkowski limit of FLRW, neglecting the curvature of spacetime or, equivalently, assuming that the time variation of the scale factor is negligible. If the time dependence cannot be neglected, the action can only be diagonalized at fixed time slices and flavour oscillation during the propagation of the phonons is unavoidable. In a weakly time dependent background, we can then diagonalize the action at a certain instant and the mixing that will happen after a short time will be proportional to $\ch$, which is the source of breaking of time translations. 

To see in some more detail the problem with the time dependence, let us neglect the metric perturbations and decompose the fields into longitudinal and transverse modes: $\pi_A  =\pi_{A\parallel}+\pi_{A\perp}$, exactly as we did in \eq{comps} for the single component case.  Using these variables, each part can be treated separately because they do not mix at quadratic order. The matter Lagrangians for longitudinal and transverse modes can be written as the sums of  quadratic forms in flavour space:
\begin{align} \label{simpleact}
\mathcal{L}_\parallel^{(2)}=\left(\dot\pi_\parallel\right)^t\cdot  {\bf X}_\parallel(\tau)\, \dot\pi_\parallel+ \left(\nabla\cdot\pi_\parallel\right)^t  {\bf Y}_\parallel(\tau)\, \nabla\cdot\pi_\parallel\quad \text{and}\quad \mathcal{L}_\perp^{(2)}=\left(\dot\pi_\perp\right)^t\cdot {\bf X}_\perp(\tau)\, \dot\pi_\perp
\end{align}
In this notation $\pi$ is a column vector in flavour space with $N$ components, $\pi^t$ is its transpose and $\bf{X}_\parallel$ etc. are real, symmetric and time dependent matrices of dimension $N\times N$. Each $\pi$ is also a (three-component) spatial vector and the dot product $\cdot$ is the scalar Euclidean product in three dimensions. If the flavour mixing matrices were constant, both Lagrangians could be diagonalized. In particular, to diagonalize the longitudinal part we would have to perform a SO(3) rotation in flavour space, a rescaling of the fields and then another rotation. Since the mixing matrices are time dependent, when we diagonalize them and rotate $\dot\pi$, we get new terms of the form $\tilde\pi^t\, {\bf M}\, \tilde\pi$ and $\tilde\pi^t\, {\bf U}\, \dot{\tilde\pi}$\, where $\tilde \pi$ is the rotated vector in flavour space and ${\bf M}$ and ${\bf U}$ are in general non-symmetric.  For small time intervals, the rate of flavour violation is controlled by the Hubble parameter $\ch$ since the mixing matrices depend on time in this form: ${\bf X}_{\parallel,\perp}={\bf X}(\ch \tau)$ and ${\bf Y}_\parallel={\bf Y}_{\parallel,\perp}(\ch\tau)$. Therefore, the mixing effects that appear after diagonalizing at a fixed time will go as $\ch$ as a first approximation.

It is important to remark the distinction between components and flavours. While the components are the different triads of $\Phi$ fields that constitute the fluid, the really important variables in the EFT of fluids are the phonons $\pi$, for which we have introduced the concept of flavour. In order to have well-defined flavours, we required diagonal kinetic terms in the phonon action, but we have just seen that it is impossible to get flavour conservation at all times in dynamic spacetimes. In the next section we are going to show that regardless of this feature of the quadratic  action \eq{seca}, we can still interpret the linear equations of motion for the phonons in terms of separable species.

\subsection{Equations of motion and taxonomy of cosmological species} \label{taxo}
As we explained in Section \ref{eftmc}, having a single energy-momentum tensor poses a problem for the definition of different species in the EFT of an $N$-component fluid. It is not apparent if and how it is possible to identify parts of the energy-momentum tensor with different species. We also saw that the action \eq{mf} for the $\Phi$ components  cannot be separated into the sum of the actions of $N$ free fluids plus interaction terms. Moreover,  the impossibility of diagonalizing the quadratic action for the phonons in a time dependent background adds up to the issue. Given all this, it would seem hopeless attempting to apply the EFT of a multi-component fluid to study common situations in cosmology where there is more than one species, such as the late time evolution of dark matter and dark energy. On the contrary, using the equations of motion, we are going to show that the EFT of a multi-component fluid is the natural framework to describe this kind of situation, and we can actually use this theory to constrain broad classes of models from the data. 

To simplify the discussion, let us think of a universe with just two species. This can be directly applied to the aforementioned dark matter -- dark energy system\footnote{The generalization to a higher number of components is immediate if we keep the analysis up to linear level in fluctuations. To describe second and higher order  cosmological perturbations with four or more species we have to include the effect of $\Psi$.}.
Concerning the background evolution, we just need to model the history of $\ch$. This can be easily done by choosing adequately $\bar F$ (which determines the background density) and  $\bar F_{J_{AB}}$, according to the equation of state of the universe \eq{eom}. In the context of dark matter and dark energy, as far as the data remains of purely gravitational nature, we can only constrain a single dark fluid. This point is emphasized e.g. in \cite{Kunz:2009yx} with a background evolution study. Unless extra theoretical assumptions are added, there exists an ambiguity (that has been  termed `dark degeneracy') in the identification of separate dark matter and dark energy species.  

We will now see how the same degeneracy appears for linear perturbations, using the EFT framework. Let us label the two components  $\Phi_1$ and $\Phi_2$. Since we will work at linear order and therefore $\eq{theta}$ has no effect, the results that we obtain below can be generalized to any number of components very easily. The longitudinal and transverse  linear equations of motion for $\pi_{1\parallel}$ and $\pi_{1\perp}$, the phonons coming from the component $\Phi_1$, are:
\begin{align}\label{complexcont}
\frac{\partial}{\partial\tau}\left(\frac{1}{a^2}\big(\left(2 \bar F_{J_{11}}+\bar F_{J_{12}}\right)\nu^i
-2\bar F_{J_{11}}\dot\pi^i_{1\perp}-\bar F_{J_{12}}\dot\pi^i_{2\perp}\big)\right)=0\\ \label{complexeuler}
\frac{\partial}{\partial\tau}\left(\frac{1}{a^2}\left(2\bar F_{J_{11}}\nabla\cdot\dot\pi_{1}+\bar F_{J_{12}}\nabla\cdot\dot\pi_{2}\right)\right)-\frac{1}{a^2}
\nabla^2\left(\mathcal{E}_{1\phi}\,\phi+\mathcal{E}_{1\psi}\,\psi+\mathcal{E}_{11}\nabla\cdot\pi_{1}+\mathcal{E}_{12}\nabla\cdot\pi_{2}\right)=0 
\end{align}
where
\begin{align}
\mathcal{E}_{1\psi}&=2\,\bar F_{J_{11}}+\,\bar F_{J_{12}}\\
\mathcal{E}_{11}&=\left(4\,\bar F_{J_{11}J_{11}}+4\,\bar F_{J_{11}J_{12}}
+\, \bar F_{J_{12}J_{12}}\right) \bar b\,^2+2\,\bar F_{J_{11}}\\
\mathcal{E}_{12}&=\left(2\, \bar F_{J_{11}J_{12}}+4\, \bar F_{J_{11}J_{22}}
+\, \bar F_{J_{12}J_{12}}+2\,\bar F_{J_{12}J_{22}}\right) \bar b\,^2+\, \bar F_{J_{12}}\\
 \mathcal{E}_{1\phi}&=\left(12\,\bar F_{J_{11}J_{11}}+18\, \bar F_{J_{11}J_{12}}+12\, \bar F_{J_{11}J_{22}}
+6\, \bar F_{J_{12}J_{12}}+6\, \bar F_{J_{12}J_{22}}\right)\bar b\,^2+6\, \bar F_{J_{11}}+3\, \bar F_{J_{12}}
\end{align}
Clearly, analogous equations hold for $\pi_2$ with the appropriate replacement of indices. The equations of motion for the phonons reflect the mixing inside the action \eq{seca}. There are multiple ways in which we can arrange the variables, defining fluid-like quantities for two species that will satisfy an Euler and a continuity equation (for the longitudinal modes) and another equation that describes the conservation of vorticity (for transverse modes). All these ways of assigning fluid variables are related through linear field redefinitions (e.g. SO(3) rotations in \eq{seca}). In the EFT of a multi-component fluid, the dark degeneracy that we mentioned earlier naturally arises as a consequence of this multiplicity of possible assignments for the density, pressure and velocity perturbations of the two species. 

Among these possible choices, there is a very convenient one that becomes nearly obvious when the entrainment is neglected, so we will first focus on that simpler case before studying the more general situation described by \eq{complexcont} and \eq{complexeuler}. As we have explained, what we aim to is interpreting the equations for propagation of phonons  in terms of fluid equations (Euler, continuity and momentum conservation) for the perturbations of coupled species. We are going to show that even if we have just one fluid and the action for its perturbations cannot be diagonalized, we can still think of the fluid as being composed by different species that interact with each other. One can easily understand why this works recalling that we remain at linear order in fluctuations. By simply Taylor expanding the energy momentum tensor we obtain that $\delta T^{\mu\nu}$ is a sum of $\delta T^{\mu\nu}_A$ terms, which in turn allow us to define separate fluid perturbation variables.

\subsubsection{Zero entrainment} \label{zero}
Let us first see in this case how to find a suitable mapping of phonon variables $\pi_A$ to fluid variables (density, pressure, etc.) corresponding to different cosmological species. Neglecting the entrainment, the phonon equations \eq{complexcont} and \eq{complexeuler} simplify to:
\begin{align} \label{vortical}
\frac{\partial}{\partial\tau}\big(a\,\bar F_{b_1}\left(\nu^i-\dot\pi_{1\perp}^i\right)\big)&=0\\ \label{longsimp}
\frac{1}{a}\frac{\partial}{\partial\tau}\big(a\,\bar F_{b_1}\nabla\cdot\dot\pi_1\big)-3\,\bar b\left(\bar F_{b_1b_1}+\bar F_{b_1b_2}\right)\nabla^2\phi-\bar F_{b_1}\nabla^2\psi-\bar b\, \bar F_{b_1b_1}\nabla^2\left(\nabla\cdot \pi_1\right)-\bar b\, \bar F_{b_1b_2}\nabla^2\left(\nabla\cdot \pi_2\right)&=0\,,
\end{align}
where we have used the following relations:
\begin{align}
{2\,b_A}F_{J_{AA}}=F_{b_{A}}\,,\quad 4\,b_A^3\,F_{J_{AA}J_{AA}}=b_A\,F_{b_{A}b_{A}}-F_{b_{A}}\\ {4\,b_Ab_B}F_{J_{AA}J_{BB}}=F_{b_Ab_B}\,,\quad A\neq B \qquad\qquad
\end{align}
to replace current contractions $J_{AA}$ by determinants $b_A$, which are the only important operators at this order when there is no entrainment. What we now want to do is to define variables that make these equations resemble the standard perturbation equations for two fluids. First, we simply split the total density and pressure perturbations of the fluid, expressions \eq{firstdenp} and \eq{secdenp}, defining: 
\begin{align} \label{encomp}
\delta\rho_A&\equiv-\bar b\, \bar F_{b_A} \left(\nabla\cdot\pi_A+3\phi\right)\\
\delta p_A &\equiv -\bar b\,^2 \left(\bar F_{b_Ab_A}+\bar F_{b_Ab_B}\right) \left(\nabla\cdot\pi_A+3\phi\right)\,,\label{encomp2}
\end{align}
in such a way that their sums give the total energy density and pressure perturbations.  Then, taking into account that the velocity divergence of each component is by construction\footnote{Recall that this expression is automatically enforced by the requirement that the rest frame four-velocity of the fluid should be constant along the flow of the fluid.}
\begin{align} \label{sctheta}
\theta_{A}=-\nabla\cdot\dot \pi_A\,,
\end{align}
it is easily checked that a continuity equation analogous to \eq{continuity} is automatically satisfied for the perturbations of each component, provided that we introduce individual equations of state via:
\begin{align} \label{sceq}
\left(1+w_A\right)\Omega_A\equiv\frac{\bar b \bar F_{b_A}}{\bar F}\,,
\end{align}
where, as usual, $\Omega_A=\bar\rho_A/\bar \rho$. It is important to remark that so far we have not used the equations of motion for the phonons and therefore \eq{continuity} holds identically. It is also worth stressing that the definition \eq{sceq} does not specify how much of $\bar\rho=-\bar F$ belongs to the energy density of each component. It only determines the sums of background density and pressure $\bar\rho_A+\bar p_A$. 

If we now use \cref{encomp,encomp2,sctheta} into \eq{longsimp} and compare the result with \eq{euler}, we obtain that the scalar anisotropic stress is given by:
\begin{align} \label{cd2simple}
\sigma_A=\frac{\bar b^2\bar F_{b_Ab_B}}{\left(\,\bar\rho_A+\bar p_A\,\right)\bar F}\left(\,\nabla\cdot\pi_A-\,\nabla\cdot\pi_B\right)\,.
\end{align}
Each species exhibits an effective anisotropic stress that depends on the difference of the divergences of the phonons. This reflects the intrinsic interacting nature of the components, which originates in the symmetry $\mathcal{V}\text{Diff}^N$ that we have used to construct the effective action. Let us point out that permuting the labels of the components on \eq{cd2simple} we flip the sign of the equality and therefore, taking the sum of the two equations we get zero. This is consistent with the fact that the total anisotropic stress of the multi-component fluid is zero.

To complete the picture, we just need the interpretation of the equations of motion of the transverse modes \eq{vortical}. This equation appears in exactly the same form in the case of a single component fluid  (or for a system of fluids that do not interact other than gravitationally) \cite{Ballesteros:2012kv} and it is a consequence of vorticity conservation. Besides, it is also the equation for the conservation of the three-momentum of each component \cite{Ballesteros:2012kv}. In the Appendix \ref{vort} we discuss the vorticity of the multi-component fluid in further detail. 

We have shown that in the zero entrainment case it is possible to define fluid-like variables \cref{encomp,encomp2,sctheta,sceq,cd2simple} that allow us to write for each component a dynamical identity which is equivalent to the continuity equation and, also, to interpret the equations of motion of the longitudinal \eq{longsimp} and transverse \eq{vortical} phonons  as Euler and momentum conservation equations, respectively. Therefore, in spite of the problems to diagonalize the action \eq{seca} and to split the energy-momentum tensor \eq{emt}, effectively, the multi-component fluid (with no entrainment) can be seen as a system of two separate (interacting) fluids with anisotropic stress, at linear order in perturbations. 

Before moving into the analysis of the form of the equations of motion in  the general case (i.e. for $J_{12}\neq 0$), let us make a comment on the time evolution of the ratios between the pressure and the density at the background and linear levels. For each component, we can define the so-called {\it adiabatic sound speed}: $\dot{\bar p}_A=c_A^2\dot{\bar \rho}_A$ using \eq{sceq}. It is simply
\begin{align} \label{sps}
c_A^2=\bar b\, \frac{\bar F_{b_Ab_A}+\bar F_{b_Ab_B}}{\bar F_{b_A}}\,.
\end{align}
By looking at the definitions \eq{encomp} and \eq{encomp2}, we see that the adiabatic sound speed coincides with the ratio between the pressure and the density perturbations, which for any fluid is commonly referred to as the {\it sound speed} of a fluid: $\delta p_A=c_{sA}^2 \delta \rho_A$. For any cosmological fluid, the adiabatic sound speed is a function of time that measures how fast the background pressure changes with respect to the background density. The sound speed $c_s^2$ measures instead the change in the pressure induced by a density perturbation and in general it can be space and time dependent. In the particular case we are studying here, our definitions  \cref{encomp,encomp2,sctheta,sceq,cd2simple} naturally lead to both speeds being equal and therefore each component behaves as an {\it adiabatic} fluid. However, it is important to point out that \eq{sps} does not correspond to the speed of sound  of propagation of $\pi_A$ waves, as it can be checked from the action \eq{seca}.

\subsubsection{General case} \label{gc}
If the entrainment cannot be neglected, it is still possible to interpret the equations of motion of the phonons in terms of distinct cosmological species. However, there is a difference with respect to the previous case because now we need to define the fluid-like variables combining the phonons of different components. The relevant equations now are \eq{complexcont} and \eq{complexeuler}. The first of them, which is the equation for the transverse modes is again a consequence of the dynamics of vorticity, which we discuss in Appendix \ref{vort}. 

The Euler equation must come from \eq{complexeuler}, whose left hand side suggests that we can combine the two phonons to define two independent velocity divergences as follows:
\begin{align} \label{double}
\begin{split} 
\left(1+w_1\right)\bar\rho_1\theta_1\equiv\bar b^2\left(2 \bar F_{J_{11}}\nabla\cdot\dot\pi_1+\bar F_{J_{12}}\nabla\cdot\dot\pi_2\right)\,,\\
\left(1+w_1\right)\bar\rho_2\theta_2\equiv\bar b^2\left(2 \bar F_{J_{22}}\nabla\cdot\dot\pi_2+\bar F_{J_{12}}\nabla\cdot\dot\pi_1\right)\,.
\end{split}
\end{align}
Although we are using the subscripts $1,2$ on both sides of these expressions, it must be clear that $\theta_1$ and $\theta_2$ are both mixtures of the two phonon components $\pi_1$ and $\pi_2$. As we are going to see immediately,  these definitions work as an effective diagonalization of the system at the level of the equations of motion. It can be easily checked that the expressions \eq{double} reduce to \eq{sctheta} by setting the entrainment to zero.   

If we also define the density and pressure perturbation of one of the species\footnote{And analogously for the other species.} to be
\begin{align} \label{dena}
\delta\rho_1&\equiv-\bar b^2\big(2\bar F_{J_{11}}\left(
\nabla\cdot\pi_1+3\phi\right)+\bar F_{J_{12}}\left(
\nabla\cdot\pi_2+3\phi\right)\big)\\
\delta p_1 &\equiv\delta\rho_1 -2\,\bar b^4\sum_{\{AB\}}\big( 2\bar F_{J_{11}J_{AB}}\left(
\nabla\cdot\pi_1+3\phi\right)+\bar F_{J_{12}J_{AB}}\left(\nabla\cdot\pi_2+3\phi\right)\big)\label{presa}
\end{align} 
 and the equation of state as
\begin{align} \label{eque}
\left(1+w_1\right)\Omega_1 &\equiv\frac{2 \bar F_{J_{11}}+\bar F_{J_{12}}}{\bar F}\,\bar b^2\,,
\end{align}
the continuity equation \eq{continuity} is identically satisfied, in the same way it happened in the zero entrainment case. Proceeding as before for the Euler equation, the anisotropic stress \eq{cd2} now takes the form
\begin{align} \label{cd3simple}
\sigma_1=\bar b^4\,\frac{4\bar F_{J_{11}J_{22}}-\bar F_{J_{12}J_{12}}}{\left(\,\bar\rho_1+\bar p_1\,\right)\bar F}\,\nabla\cdot\left(\pi_1-\pi_2\right)\,,
\end{align}
which reduces to \eq{cd2simple} for zero entrainment.  

The adiabatic sound speed can be defined for each component using \eq{eque}, exactly as we did in the case with no entrainment. However, it is now more difficult  to define a sound speed for the perturbations, due to the different $\pi$ dependencies of the density and the pressure perturbations,  \eq{dena} and \eq{presa} respectively. If we insist in using the ratio between the two, we get a fraction of $\mathcal{O}(1)$ quantities in $\pi$. So, when the entrainment is non-negligible, the fluid variables we have defined with \cref{double,dena,presa,eque} are {\it non-adiabatic}, because $\dot {\bar p}_A \delta\rho_A\neq \dot {\bar \rho}_A \delta p_A$\,.

\subsubsection{From anisotropic stress to interacting species} \label{inter}
The results obtained in Sections \ref{zero} and \ref{gc} provide the simplest maps that allow to interpret the propagation of the phonons in terms of the fluid variables of a system of different species. As we have seen, these species naturally turn out to have anisotropic stresses: \eq{cd2simple} and \eq{cd3simple} in the cases without and with entrainment, respectively. The dependence of the anisotropic stresses on phonons coming from different components explicitly shows the interacting nature of the system and the impossibility of diagonalizing the quadratic action at all times.

 Although those maps of variables are very convenient, they are not unique. We are going to see now that the phonon equations can actually be reinterpreted in terms of couplings $\mathcal{Q}^\mu_A$ between species, while still keeping the definitions for the energy density, pressure and velocity fluctuations. The results that we present in this section show that in the EFT of a multi-component fluid a coupling between species at linear order in perturbations can always be recast into anisotropic stresses through a simple redefinition of variables.

In general, in a system whose energy momentum tensor can be written as a sum of tensors for different species
\begin{align}
T^{\mu\nu}=\sum_A T^{\mu\nu}_A
\end{align}
each $T^{\mu\nu}_A$ is not individually conserved and we write:
\begin{align} \label{source}
T^{\mu\nu}_{A\,\,\,;\mu}=\mathcal{Q}^\nu_A\,,\quad \sum_{A=1}^N\mathcal{Q}^\nu_A=0.
\end{align}
Each source term can be decomposed as follows\footnote{Kodama and Sasaki \cite{Kodama01011984} decompose the source $\mathcal{Q}^A_{\mu}$ into a component parallel to the rest frame of the total fluid  (which is given by \eq{trf} in our case) and a vector orthogonal to it. We prefer to project onto the velocities of the components themselves and their orthogonal directions.}:
\begin{align}
\mathcal{Q}^A_{\mu}=Q_A u^A_\mu+\ch(\rho_A+p_A)\, \mathcal{U}^A_\mu\,,\quad g^{\mu\nu}u_\mu^A\, \mathcal{U}_\nu^A=0\,,
\end{align}
where 
\begin{align}
Q_A=\bar{Q}_A+\delta Q_A\,,\quad \mathcal{\bar U}_0^A=0\,,\quad \mathcal{U}^A_i=(\bar \rho_A+\bar p_A)\ch U^A_i
\end{align}
and ${U}^A_i$ is a first order quantity in cosmological perturbation theory. The continuity and Euler  equations, \eq{continuity} and \eq{euler}, of each species are modified by the source $\mathcal{Q}^\mu_A$:
\begin{align}
\label{continuityQ}
\frac{\partial}{\partial\tau}\left(\delta\rho_A\right) &=-3\ch\left(\delta\rho_A+\delta p_A\right)+\left(\bar\rho_A+\bar p_A \right)\left(3\dot\phi-\theta_A\right)+a\,\delta Q_a-a\,\bar Q_A \psi\\
\nonumber
\bar\rho_A\,\frac{\partial}{\partial\tau}\big((1+w_A)\,\theta_A\big)&=(3w_A-1)\left(\,\bar\rho_A+\bar p_A\,\right)\ch\theta_A -\nabla^2 \delta p_A+\left(\,\bar\rho_A+\bar p_A\,\right)\nabla^2(\sigma_A-\psi)\\&+a\,\bar Q_A \theta_A+a\ch\left(\,\bar\rho_A+\bar p_A\,\right)\nabla\cdot U_A\,.\label{eulerQ}
\end{align}

As we discussed earlier, it is not apparent how to decompose the energy-momentum tensor of the effective multi-component fluid \eq{emt} into a sum of tensors, but in practice we can formally do it at linear order and this allows us to interpret the longitudinal phonon equations of motion  with the formalism we just described, using \eq{continuityQ} and \eq{eulerQ}. In the zero entrainment case, comparing \eq{continuity} and \eq{continuityQ} we find that the assignment of variables \cref{encomp,encomp2,sctheta,sceq} implies the following constrain on $\mathcal{Q}^\mu_A$:
\begin{align} \label{cd1}
\delta Q_A=\bar Q_A \psi\,,
\end{align}
because \eq{continuity} is identically satisfied. Besides, using \eq{eulerQ} we find that \eq{cd2simple} must be replaced by
\begin{align} \label{cd2}
\frac{\bar b^2\bar F_{b_Ab_B}}{\bar F}\,\nabla^2\left(\,\nabla\cdot\pi_A-\,\nabla\cdot\pi_B\right)=\left(\,\bar\rho_A+\bar p_A\,\right)\nabla^2\sigma_A+a\ch\,\left(\,\bar\rho_A+\bar p_A\,\right)\,\nabla\cdot U_A+a\,\bar Q_A \theta_A\,.
\end{align}
The system of equations \eq{cd1} and \eq{cd2} have various solutions, each of which leads to different interpretations of our multi-component fluid. One possibility is to set $\mathcal{Q}^\mu_1=-\mathcal{Q}^\mu_2=0$ as we did before, leading to \eq{cd2simple}.  Conversely, we can also have $\mathcal{Q}^{\mu}_1\neq 0$ and set the anisotropic stresses to zero. This trade-off between the source terms $\mathcal{Q}^\mu_A$ and the anisotropic stresses is an interesting property. At the root of it lies the fact that the effective anisotropic stress of each component arises due to its interaction with the other component. 
Analogously, if we allow the entrainment to be different from zero, the condition \eq{cd1} still holds and the analogous of the expression \eq{cd3simple} is
\begin{align} \label{cd3}
\bar b^4\,\frac{4 \bar F_{J_{11}J_{22}}- \bar F_{J_{12}J_{12}}}{\bar F}\,\nabla^2\,\left(\nabla\cdot\pi_1-\nabla\cdot\pi_2\right)=\left(\,\bar\rho_1+\bar p_1\,\right)\nabla^2\sigma_1+a\ch\,\left(\,\bar\rho_1+\bar p_1\,\right)\,\nabla\cdot U_1+a\,\bar Q_1 \theta_1\,.
\end{align}
If we think in terms of the dark matter -- dark energy example, these results tell us that an anisotropic stress can be mimicked by (what in cosmology is usually called) a fluid interaction and both have the same physical origin.  As we already mentioned before, the `dark degeneracy' is just a consequence of the interacting nature of the theory. 

Let us recall  that the source $\mathcal{Q}^\mu_A$ also affects the relation between the sound speeds of each species. In particular,  we saw at the end of the Section \ref{zero} that the adiabatic and non-adiabatic sound speeds of each component are equal if the entrainment is zero. However, that result was obtained assuming that $Q^\mu_A$ is zero. In the more general case that we are now considering, the time variation of the background density of each component is 
\begin{align}
\dot{\bar\rho}_A=-3\ch\left(\bar\rho_A+\bar p_A\right)+\bar Q_A\,,
\end{align}
introducing a difference between the two speeds:
\begin{align}
1+c_{sA}^2=\left(1+c_A^2\right)\left(1+\frac{\bar Q_A}{3\ch(\bar\rho_A+\bar p_A)}\right)\,,
\end{align}
where $c_A^2$ is still given by \eq{sps}.

\section{\,\,Summary and outlook} \label{conc}

In this paper we presented the EFT of multi-component fluids. Our motivation has been the development of a formalism for describing common situations in which different cosmological species may interact not only through gravity. 

We have identified the operators that contribute to the effective action for the phonons at the lowest order in derivatives. These operators are determined by symmetry requirements on the comoving coordinates of the components $\Phi_A$. Specifically, we impose invariance under $\mathcal{V}\text{Diff}^N$, which is the most natural extension of the single-component perfect fluid case. The invariants are of three kinds and they can all be written in terms of the covariantly conserved currents $J_A^\mu$. First, there are the determinants $b_A=J_{AA}$,  that are analogous variables to the entropy density in a single-component case. Then, we have the entrainments $J_{AB}$ with $A\neq B$ for different components. And finally, we have also found a new type of operator, $\Psi$, which has the interesting property of behaving as a cosmological constant if the Lagrangian depends exclusively on it. On the contrary, if the operators $J_{AB}$ are also taken into account, $\Psi$ is irrelevant at the background level and its effect can only be seen on the perturbations. The expression of $\Psi$ in terms of phonons is given in \eq{psi}. This structure could  give an interesting non-Gaussian shape in the three-point function of velocity correlators of different  species. It could also be important in a model of multi-field inflation that would generalize \cite{Endlich:2012pz}. We leave a study of the cosmological implications of $\Psi$ for future work. 

Since the fully non-linear Lagrangian \eq{mf} is a general function of all the operators, we can only define a single gravitational energy-momentum tensor for the entire system. This is the reason why we refer to it as a multi-component fluid instead of as an ensemble of several fluids. We have also discussed the notion of {\it flavour}, which we have identified as the excitations of the multi-component fluid that can propagate independently. In other words, the flavours are defined by the degrees of freedom that diagonalize the quadratic action for the phonons. 

We have shown that in  time dependent backgrounds (which typically appear in cosmological applications) the quadratic phonon action cannot be diagonalized at all times with conserved flavours. The diagonalization is only possible at slices of constant time and then, as time flows, unavoidable flavour mixing takes place. We argued that the amount of flavour violation for short time intervals is proportional to the Hubble parameter $\ch$ and therefore it can be neglected for propagation times much shorter than the age of the universe, or in situations in which a Minkowski approximation for the metric is adequate.

In spite of the non-diagonalizability of the quadratic phonon action, a multi-component effective fluid can be interpreted in terms of separate cosmological species at the level of the linear equations of motion for the perturbations. We have shown this explicitly by choosing the adequate definitions for the density, pressure and anisotropic stress of the species. There are different mappings that are possible. In the simplest one of them, we effectively decompose the total energy momentum tensor of the multi-component fluids into the tensors of species that are separately conserved. In this case, the interaction of the phonons manifests itself into anisotropic stress terms that depend on phonons coming from different components. 

We also considered a more general possibility in which only the total energy-momentum tensor of the multi-component fluid is covariantly conserved. We have shown that in this case the anisotropic stresses can be recast into sources for the energy momentum-tensors of the individual species through a redefinition of variables. 

Our results can be applied to a wide variety of systems in cosmology. As an example we have mentioned that the ambiguity in defining dark matter and dark energy, called dark degeneracy, can be thought to originate (in the context of this effective field theory) in the phonon mixing; and it implies the possibility of interpreting the equations of motion in different ways.

An attractive direction to extend this work would be to change the symmetry $\mathcal{V}\text{Diff}^N$. For example, we could have simply imposed invariance under translations and $SO(3)$ rotations of each component. This would produce a more complex fluid, having perhaps extra features that it may be interesting to explore, and it would correspond to the direct generalization of the single-component `solid' discussed in detail in \cite{Endlich:2012pz}.  
Finally, the thermodynamics of the multi-component fluid remains to be solved. Specifically it would be necessary to find the appropriate definitions of entropy, temperature and chemical potentials.

\appendix
\renewcommand\thesection{A\arabic{section}}

\section{\,\,Vorticity} \label{vort}

We can define vorticity currents and charges in the same way as for a single component fluid \cite{Dubovsky:2005xd,Ballesteros:2012kv}. The symmetry \eq{difflavour} of each component gives rise to a conserved vorticity current via Noether theorem.
If there is no entrainment and there is no $\Psi$, the results for the vorticity are very similar to the ones valid for a single fluid. Under those conditions, any current of the form
\begin{align} \label{vortc}
\mathcal{J}^\mu_{A}=-b_A F_{b_A} \epsilon^i\left(\Phi_A\right)\left(B^{-1}_A\right)^{ij}\partial^\mu\Phi^j_A
\end{align}
is conserved provided that $\partial \epsilon^i/\partial \Phi_A^i=0$, which is the condition for volume conservation of the infinitesimal  internal diffeomorphisms  of the $A$--th component: $\Phi^i_A\rightarrow\Phi^i_A+\epsilon^i(\Phi_A)$. This leads to an infinite set of conserved charges\footnote{In spite of the notation, these charges $Q$ have nothing to do with the sources of the energy momentum tensors in Section \ref{inter}.}:
\begin{align} \label{chargef}
Q^{A}_f=\int d^3 \Phi_A\, f_a(\Phi_A)\,Q^{A}_a\,,
\end{align}
where
\begin{align} \label{efe}
\epsilon^i(\Phi_A)=\epsilon_{ijk}\partial_jf_k\left(\Phi_A\right)
\end{align}
and 
\begin{align}
Q^{A}_a=\epsilon_{abc}\partial_b V_{c}^{A}\,,\quad V_{a}^{A}=-F_{b_A}\frac{\partial x^j}{\partial \Phi_A^a}u^{A}_{j}\,.
\end{align}
The quantities $Q^{A}_a$ are the vorticity charges and $V_{a}^{A}$ is related to the vorticity circulation over a surface with boundary $\partial \Sigma$: 
\begin{align}
\label{vorticitydefmetric}
\mathcal{V}  \equiv  \int_{\Sigma} \star (Q^{A}_a d\Phi_A^a)=\oint_{\partial\Sigma} V_{a}^{A} d\Phi^a_A=-\oint_{x(\tau,\partial\Sigma)} F_{b_A} u_j^{A} dx^j\,,
\end{align}
which is also conserved on-shell. For more details on the construction of the vorticity charges and the circulation, we refer the reader to \cite{Ballesteros:2012kv}, where this is done using the ADM formalism. 

The conservation of the vorticity charges $Q^{A}_a$ determines the time evolution of the transverse modes. In particular, at linear order without entrainment we obtain:
\begin{align} \label{vecttrans}
\frac{\partial}{\partial\tau}\left(a\, F_{b_A}\left(\nu^k-\dot\pi^k_{A\perp}\right) \right)=0\,,
\end{align}
which is precisely the equation \eq{vortical}. As it is explained in \cite{Ballesteros:2012kv}, this equation describes the time evolution of the three-momentum of each component, which is $\delta q^i_A\equiv(\bar\rho_A+\bar p_A)(\nu^i-\dot\pi^i_{A\perp})$\,, consistently with the definition \eq{sceq} of the sum of density and background pressure.
\subsection{Including entrainment}
We now want to compute the vorticity charges including the effect of entrainment (but still neglecting the $\Psi$ operators). In order to do it we start from the general expression \cite{Ballesteros:2012kv}:
\begin{align}
Q_{(\epsilon)}=\int d^3x\sqrt{-g} \,\mathcal{J}^0_{A(\varepsilon)} =\int d^3 \Phi_A \frac{\mathcal{J}_{A(\epsilon)}^0}{J^0_A}
\end{align}
for the conserved charge associated to an infinitesimal $\mathcal{V}\text{Diff}_A$ \eq{difflavour} parametrized by $\epsilon^i(\Phi_A)$ as above. In this expression, the quantity in the numerator of the integrand is the time component of the generalization of the current \eq{vortc} to the case of non-zero entrainment. This current has the expression
\begin{align} \label{curr1}
\mathcal{J}^\mu_{A(\epsilon)}=\epsilon^i(\phi_A)\sum_{\{BC\}}F_{J_{BC}}\frac{\partial J_{BC}}{\partial\,\partial_\mu\Phi^i_A}=\epsilon^a(\Phi_A)\sum_{B=A}^N F_{J_{AB}} S_\alpha^B M^{\mu\alpha}_{A\,a}+\epsilon^a(\Phi_A)\sum_{B=1}^A F_{J_{BA}} S_\alpha^B M^{\mu\alpha}_{A\,a}
\end{align}
where
\begin{align}
M^{\mu\alpha}_{A\,a}=-\frac{\epsilon_{abc}}{2\sqrt{-g}}\epsilon^{\mu\alpha\beta\gamma}\partial_\beta\Phi^b_A\partial_\gamma\Phi^c_A\,.
\end{align}
If we express $\epsilon(\Phi_A)$ as in \eq{efe}, we obtain an expression analogous to \eq{chargef} where
\begin{align} \label{vortq}
Q^{A}_a=\epsilon_{abc}\partial_b V_{c}^{A}\,,\quad V_{c}^{A}=\frac{1}{a^2} \sum_{B=1}^A \bar F_{J_{BA}}\left(\nu^c+v^c_B\right)+\frac{1}{a^2} \sum_{B=A}^N \bar F_{J_{AB}}\left(\nu^c+v^c_B\right)+\mathcal{O}(2)
\end{align}
Again, the conservation of the vorticity charges gives the dynamics of the transverse modes at linear order. In particular, if there are just two components, the (linearized) equation $\dot Q^A_a=0$ is the same as \eq{complexeuler}. Notice that at linear order in perturbations, the entrainment does not contribute to the vorticity charges, and therefore \eq{vortq} is completely general at this order. 

\section{\,\,Eckart frames}
Instead of projecting the energy momentum-tensor on the total energy frame, as we have done in most of this work, another possibility is selecting the rest frame \eq{rfa} of one of the components of the fluid, e.g. $u^\mu_C$, defined from \eq{rfa}. The advantage of this choice is that we know exactly the frame at all orders and therefore we can get the corresponding energy density and pressure fully non-linearly:
\begin{align} \label{ede}
\rho_{[C]}&=-F+3\,\Psi\, F_{\Psi}+2\sum_{\{AB\}}^N F_{J_{AB}}\left(J_{AB}-\frac{J_{AC}\,J_{BC}}{b_C\,^2}\right)\\
p_{[C]}&=F-3\,\Psi \, F_{\Psi}-\frac{2}{3}\sum_{\{AB\}}^N F_{J_{AB}}\left(2\,J_{AB}+\frac{J_{AC}\,J_{BC}}{b_C\,^2}\right)\,. \label{pre}
\end{align}
This can be useful to formulate the dynamics using the covariant approach to cosmological perturbation theory \cite{Ellis:1989jt}. The energy flux in such a frame can also be computed easily and is different from zero because the four-velocity $u^\mu_{C}$ is not parallel to the energy frame of the fluid:
\begin{align}
q_\mu^{[C]}=2\sum_{\{AB\}}^N \frac{F_{J_{AB}}}{b_C}\left(J^{C(A}\,J^{B)}_\mu-\frac{J_{AC}\,J_{BC}}{b_C\,^2}\,J_\mu^{C}\right)\,.
\end{align}
The anisotropic stress $\pi_{[C]}^{\mu\nu}$ is also non-zero in any component frame and can be obtained inserting the previous results into \eq{emtg}. 

\section*{\,\,Acknowledgements}

G.B. is supported by DFG through the project TRR33 ``The Dark Universe'' and thanks the hospitality of the CERN Theory Division while part of this work was done. B.B. is supported in part by the MIUR-FIRB grant RBFR12H1MW, and by the Agence Nationale de la Recherche under contract ANR 2010 BLANC 0413 01. L.M. is supported by a grant from the Swiss National Science Foundation. We thank Jose Beltr\'{a}n, Enea Di Dio, Valeria Pettorino and Ignacy Sawicki for interesting conversations.

\bibliography{bibbbm}
\bibliographystyle{hunsrt}

\end{document}